\documentclass[english]{article}
\usepackage[T1]{fontenc}
\usepackage[latin9]{inputenc}
\usepackage{amsmath}
\usepackage{amssymb}
\usepackage{esint}

\makeatletter
\newcommand{\lyxaddress}[1]{
	\par {\raggedright #1
	\vspace{1.4em}
	\noindent\par}
}

\usepackage{babel}
\date{}

\makeatother

\usepackage{babel}
\begin{document}
\title{Vacuum and spacetime signature in the theory of superalgebraic spinors}
\author{Vadim Monakhov}
\maketitle

\lyxaddress{Saint Petersburg University, Institute of Physics, Ulyanovskaya 1,
Stariy Petergof, Saint Petersburg, 198504, Russia; v.v.monahov@spbu.ru}
\begin{abstract}
We investigated action of operator analogs of Dirac gamma matrices
(we called them gamma operators) on a vacuum. We derived formulas
for vacuum state vector and operators of the Lorentz transformations
of spinors in the superalgebraic representation of spinors. Five operator
analogs of five Dirac gamma matrices exist in the superalgebraic approach
as well as two additional operator analogs of gamma matrices. Gamma
operators are constructed from Grassmann densities and derivatives
with respect to them. We have shown that there are operators which
are built from creation and annihilation operators, and that they
are also analogs of Dirac gamma matrices. However, unlike gamma operators
of the first kind, they are Lorentz invariant. We have shown that
the condition for the existence of spinor vacuum imposes restrictions
on possible variants of Clifford algebras of gamma operators: only
real algebra with one timelike basis Clifford vector corresponding
to the zero gamma matrix in the Dirac representation can be realized.
In this case, the signature of the four-dimensional spacetime, in
which there is a vacuum state, can only be (1, -1, -1, -1), and there
are two additional axes corresponding to the inner space of the spinor,
with a signature (-1, -1).
\end{abstract}

\section{Introduction}

The question of the origin of the dimension and the spacetime signature
has long attracted the attention of physicists. At the same time,
there are different approaches in attempts to substantiate the observed
dimension and the spacetime signature.

One of the main directions is the theory of supergravity. It was shown
in \cite{key-1} that the maximum dimension of spacetime, at which
supergravity can be built, is equal to 11. At the same time, multiplets
of matter fields for supersymmetric Yang-Millss theories exist only
when the dimension of spacetime is less than or equal to 10 \cite{key-2}.

Subsequently, the main attention was paid to the theory of superstrings
and supermembranes. Various versions of these theories were combined
into an 11-dimensional M-theory \cite{key-3,key-4}. In \cite{key-5},
the most general properties of the theories of supersymmetry and supergravity
in spaces of various dimensions and signatures were analyzed. Proceeding
from the possibility of the existence of majoram and pseudo-Maioran
spinors in such spaces, it was shown that supersymmetry and supergravity
of M-theory can exist in 11-dimensional and 10-dimensional spaces
with arbitrary signatures, although depending on the signature the
theory type will differ. Later, other possibilities were shown for
constructing variants of M-theories in spaces of different signatures
\cite{key-6}.

Another approaches are Kaluza-Klein theories. For example, in \cite{key-7}
it was shown that in the theories of Kaluza-Klein in some cases it
is possible not to postulate, but to determine from the dynamics not
only the dimension of the spacetime, but also its signature.

In \cite{key-8,key-9,key-10}, an attempt was made to find a signature
based on the average value of the quantum fluctuating metric of spacetime.

An attempt was made in \cite{key-11} to explain the dimension and
signature of spacetime from the anthropic principle and the possibility
of causality, in \cite{key-12} from the existence of equations of
motion for fermions and bosons coinciding with four-dimensional ones,
in \cite{key-13} from the possibility of existence in spacetime classical
electromagnetism.

In all the above approaches, the fermion vacuum operator in the second
quantization formalism is not constructed and the restrictions imposed
by such a construction are not considered. Therefore, the possibility
of the existence of a vacuum and fermions is not discussed. In particular,
the vacuum should be a Lorentz scalar and have zero spin, but in the
theory of algebraic spinors, which more generally describes spinors
than the Dirac matrix theory, Clifford vacuum has the transformational
properties of the spinor component, and not the scalar \cite{key-14}.

The author develops an approach to the theory of spacetime, allowing
to solve this problem. It is based on the theory of superalgebraic
spinors -- an extension of the theory of algebraic spinors, in which
the generators of Clifford algebras (Dirac gamma matrices) are composite.

In \cite{key-15,key-16}, it was shown that using Grassmann variables
and derivatives with respect to them, one can construct an analog
of matrix algebra, including analogs of matrix columns of 4-spinors
and their adjoint rows of conjugate spinors. But at the same time,
the spinors and their conjugates exist in the same space -- in the
same algebra.

In \cite{key-17,key-18}, this approach was developed -- Grassmann
densities $\theta^{a}(p),\,a=1,2,3,4$, and derivatives $\frac{\partial}{\partial\theta^{a}(p)}$with
respect to them were introduced, with CAR-algebra 
\begin{equation}
\{\frac{\partial}{\partial\theta^{i}(p)},\theta^{k}(p')\}=\delta(p-p')\delta_{i}^{k}\,.\label{eq:x0}
\end{equation}

Superalgebraic analogs $\hat{\gamma}^{\mu}$ (\ref{eq:x1}) are constructed
for Dirac gamma matrices $\gamma^{\mu}$ from these densities, we
call them gamma operators. 
\begin{equation}
\begin{split}\hat{\gamma}^{0} & =\int d^{3}p\,[\frac{\partial}{\partial\theta^{1}(p)}\theta^{1}(p)+\frac{\partial}{\partial\theta^{2}(p)}\theta^{2}(p)+\frac{\partial}{\partial\theta^{3}(p)}\theta^{3}(p)+\frac{\partial}{\partial\theta^{4}(p)}\theta^{4}(p),*]\,,\\
\hat{\gamma}^{1} & =\int d^{3}p\,[\frac{\partial}{\partial\theta^{1}(p)}\frac{\partial}{\partial\theta^{4}(p)}-\theta^{4}(p)\theta^{1}(p)+\frac{\partial}{\partial\theta^{2}(p)}\frac{\partial}{\partial\theta^{3}(p)}-\theta^{3}(p)\theta^{2}(p),*]\,,\\
\hat{\gamma}^{2} & =i\int d^{3}p\,[-\frac{\partial}{\partial\theta^{1}(p)}\frac{\partial}{\partial\theta^{4}(p)}-\theta^{4}(p)\theta^{1}(p)+\frac{\partial}{\partial\theta^{2}(p)}\frac{\partial}{\partial\theta^{3}(p)}+\theta^{3}(p)\theta^{2}(p),*]\,,\\
\hat{\gamma}^{3} & =\int d^{3}p\,[\frac{\partial}{\partial\theta^{1}(p)}\frac{\partial}{\partial\theta^{3}(p)}-\theta^{3}(p)\theta^{1}(p)-\frac{\partial}{\partial\theta^{2}(p)}\frac{\partial}{\partial\theta^{4}(p)}+\theta^{4}(p)\theta^{2}(p),*]\,,\\
\hat{\gamma}^{4} & =i\hat{\gamma}^{5}=i\int d^{3}p\,[\frac{\partial}{\partial\theta^{1}(p)}\frac{\partial}{\partial\theta^{3}(p)}+\theta^{3}(p)\theta^{1}(p)+\frac{\partial}{\partial\theta^{2}(p)}\frac{\partial}{\partial\theta^{4}(p)}+\theta^{4}(p)\theta^{2}(p),*]\,.
\end{split}
\label{eq:x1}
\end{equation}

They convert and their linear combinations in the same way that Dirac
matrices convert matrix columns and their linear combinations. The
theory is automatically secondarily quantized and does not require
normalization of operators.

In the proposed theory, in addition to analogs of the Dirac matrices,
there are two additional gamma operators and , the rotation operator
in whose plane (gauge transformation) is analogous to the charge operator
of the second quantization method \cite{key-18}: 
\begin{equation}
\begin{split}\hat{\gamma}^{6} & =i\int d^{3}p\,[\frac{\partial}{\partial\theta^{1}(p)}\frac{\partial}{\partial\theta^{2}(p)}+\theta^{2}(p)\theta^{1}(p)-\frac{\partial}{\partial\theta^{3}(p)}\frac{\partial}{\partial\theta^{4}(p)}-\theta^{4}(p)\theta^{3}(p),*]\,,\\
\hat{\gamma}^{7} & =\int d^{3}p\,[\frac{\partial}{\partial\theta^{1}(p)}\frac{\partial}{\partial\theta^{2}(p)}+-\theta(p)\theta^{1}(p)+\frac{\partial}{\partial\theta^{3}(p)}\frac{\partial}{\partial\theta^{4}(p)}-\theta^{4}(p)\theta^{3}(p),*]\,,
\end{split}
\label{eq:x2}
\end{equation}

In \cite{key-18}, it was shown that transformations of densities
and , while maintaining their CAR-algebra of creation and annihilation
operators, provide transformations of field operators of the form
, 
\begin{equation}
\varPsi'=(1+i\hat{\gamma}^{a}d\omega_{a}+\frac{1}{4}\hat{\gamma}^{ab}d\omega_{ab})\varPsi\,,\label{eq:x3}
\end{equation}
where $\hat{\gamma}^{ab}=\frac{1}{2}(\hat{\gamma}^{a}\hat{\gamma}^{b}-\hat{\gamma}^{b}\hat{\gamma}^{a});\,a,b=0,1,2,3,4,6,7$,
and $d\omega_{ab}=-d\omega_{ba}$ -- real infinitesimal transformation
parameters. The multiplier 1/4 is added in (\ref{eq:x3}) compared
to \cite{key-18} to correspond to the usual transformation formulas
for spinors in the case of Lorentz transformations.

\section{Operators of pseudo-orthogonal rotation }

Operators $\hat{\gamma}^{ab}$ are generators of pseudo-orthogonal
rotations of the form $exp(\hat{\gamma}^{ab}\omega_{ab}/4)$, where$a,b=0,1,2,3,4,6,7$.
We will call them gamma operators of rotations. They are generators
of Lorentz rotations when $a,b=0,1,2,3$.

Operators of annihilation of spinors $b_{\alpha}(p),\,\alpha=1,2$,
and of antispinors $b_{\tau}(p),\,\tau=3,4$, are obtained by Lorentz
rotations from $\frac{\partial}{\partial\theta^{\alpha}(0)}$ and
$\frac{\partial}{\partial\theta^{\tau}(0)}$ , and the Dirac conjugated
to them operators of creation $\bar{b}_{\alpha}(p)$ and $\bar{b}_{\tau}(p)$
-- by Lorentz rotations from $\theta^{\alpha}(0)$ and $\theta^{\tau}(0)$
\cite{key-17,key-18}, while momentum in the argument is replaced
from 0 to $p$: 
\begin{equation}
\begin{split}b_{i}(p) & =(e^{\hat{\gamma}^{0k}\omega_{0k}/2}\frac{\partial}{\partial\theta^{i}(0)})|_{0\rightarrow p}\,,\\
\bar{b}_{i}(p) & =(e^{\hat{\gamma}^{0k}\omega_{0k}/2}\theta^{i}(0))|_{0\rightarrow p}\,,\\
i & =1,2,3,4.
\end{split}
\label{eq:x4}
\end{equation}

Anticommutation relations for $b_{i}(p)$ and $\bar{b}_{k}(p')$ 
\begin{equation}
\{b_{i}(p),\bar{b}_{k}(p')\}=\delta(p-p')\delta_{i}^{k}\,.\label{eq:x5}
\end{equation}

In (\ref{eq:x4}), the particle momentum $p$ depends on Lorentz rotation
parameters $\omega_{0k}$. For example, for rotation in the plane
$\hat{\gamma}^{0},\hat{\,\gamma}^{1}$ the transformation (4) for
$b_{1}(p)$ and $\bar{b}_{1}(p)$ will look like 
\begin{equation}
\begin{split}b_{1}(p) & =\cosh\frac{\omega_{01}}{2}\frac{\partial}{\partial\theta^{1}(p)}+\sinh\frac{\omega_{01}}{2}\hat{\gamma}^{01}\frac{\partial}{\partial\theta^{1}(p)}\,,\\
\bar{b}_{1}(p) & =\cosh\frac{\omega_{01}}{2}\theta^{1}(p)+\sinh\frac{\omega_{01}}{2}\hat{\gamma}^{01}\theta^{1}(p)\,.
\end{split}
\label{eq:x6a}
\end{equation}

As a result, we get 
\begin{equation}
\begin{split}b_{1}(p) & =\cosh\frac{\omega_{01}}{2}\frac{\partial}{\partial\theta^{1}(p)}+\sinh\frac{\omega_{01}}{2}\theta^{4}(p)\,,\\
\bar{b}_{1}(p) & =\cosh\frac{\omega_{01}}{2}\theta^{1}(p)-\sinh\frac{\omega_{01}}{2}\frac{\partial}{\partial\theta^{4}(p)}\,.
\end{split}
\label{eq:x6}
\end{equation}

Expressions for operators $\hat{\gamma}^{ab}$ are given in (\ref{eq:x7a})
-- they will be important later. 
\begin{equation}
\begin{split}\hat{\gamma}^{01} & =\int d^{3}p\,[\frac{\partial}{\partial\theta^{1}(p)}\frac{\partial}{\partial\theta^{4}(p)}+\theta^{4}(p)\theta^{1}(p)+\frac{\partial}{\partial\theta^{2}(p)}\frac{\partial}{\partial\theta^{3}(p)}+\theta^{3}(p)\theta^{2}(p),*]\,,\\
\hat{\gamma}^{02} & =-i\int d^{3}p\,[\frac{\partial}{\partial\theta^{1}(p)}\frac{\partial}{\partial\theta^{4}(p)}-\theta^{4}(p)\theta^{1}(p)-\frac{\partial}{\partial\theta^{2}(p)}\frac{\partial}{\partial\theta^{3}(p)}+\theta^{3}(p)\theta^{2}(p),*]\,,\\
\hat{\gamma}^{03} & =\int d^{3}p\,[\frac{\partial}{\partial\theta^{1}(p)}\frac{\partial}{\partial\theta^{3}(p)}+\theta^{3}(p)\theta^{1}(p)-\frac{\partial}{\partial\theta^{2}(p)}\frac{\partial}{\partial\theta^{4}(p)}-\theta^{4}(p)\theta^{2}(p),*]\,,\\
\hat{\gamma}^{04} & =i\int d^{3}p\,[\frac{\partial}{\partial\theta^{1}(p)}\frac{\partial}{\partial\theta^{3}(p)}-\theta^{3}(p)\theta^{1}(p)+\frac{\partial}{\partial\theta^{2}(p)}\frac{\partial}{\partial\theta^{4}(p)}-\theta^{4}(p)\theta^{2}(p),*]\,,\\
\hat{\gamma}^{06} & =i\int d^{3}p\,[\frac{\partial}{\partial\theta^{1}(p)}\frac{\partial}{\partial\theta^{2}(p)}+\theta^{2}(p)\theta^{1}(p)-\frac{\partial}{\partial\theta^{3}(p)}\frac{\partial}{\partial\theta^{4}(p)}+\theta^{4}(p)\theta^{3}(p),*]\,,\\
\hat{\gamma}^{07} & =\int d^{3}p\,[\frac{\partial}{\partial\theta^{1}(p)}\frac{\partial}{\partial\theta^{2}(p)}+-\theta(p)\theta^{1}(p)+\frac{\partial}{\partial\theta^{3}(p)}\frac{\partial}{\partial\theta^{4}(p)}+\theta^{4}(p)\theta^{3}(p),*]\,,\\
\hat{\gamma}^{12} & =-i\int d^{3}p\,[\frac{\partial}{\partial\theta^{1}(p)}\theta^{1}(p)-\frac{\partial}{\partial\theta^{2}(p)}\theta^{2}(p)-\frac{\partial}{\partial\theta^{3}(p)}\theta^{3}(p)+\frac{\partial}{\partial\theta^{4}(p)}\theta^{4}(p),*]\,,\\
\hat{\gamma}^{13} & =\int d^{3}p\,[\frac{\partial}{\partial\theta^{1}(p)}\theta^{2}(p)-\frac{\partial}{\partial\theta^{2}(p)}\theta^{1}(p)+\frac{\partial}{\partial\theta^{3}(p)}\theta^{4}(p)-\frac{\partial}{\partial\theta^{4}(p)}\theta^{3}(p),*]\,,\\
\hat{\gamma}^{14} & =i\int d^{3}p\,[\frac{\partial}{\partial\theta^{1}(p)}\theta^{2}(p)+\frac{\partial}{\partial\theta^{2}(p)}\theta^{1}(p)+\frac{\partial}{\partial\theta^{3}(p)}\theta^{4}(p)+\frac{\partial}{\partial\theta^{4}(p)}\theta^{3}(p),*]\,,\\
\hat{\gamma}^{16} & =i\int d^{3}p\,[-\frac{\partial}{\partial\theta^{3}(p)}\theta^{1}(p)-\frac{\partial}{\partial\theta^{1}(p)}\theta^{3}(p)+\frac{\partial}{\partial\theta^{4}(p)}\theta^{2}(p)+\frac{\partial}{\partial\theta^{2}(p)}\theta^{4}(p),*]\,,\\
\hat{\gamma}^{17} & =\int d^{3}p\,[\frac{\partial}{\partial\theta^{3}(p)}\theta^{1}(p)-\frac{\partial}{\partial\theta^{1}(p)}\theta^{3}(p)-\frac{\partial}{\partial\theta^{4}(p)}\theta^{2}(p)+\frac{\partial}{\partial\theta^{2}(p)}\theta^{4}(p),*]\,,\\
\hat{\gamma}^{23} & =-i\int d^{3}p\,[\frac{\partial}{\partial\theta^{1}(p)}\theta^{2}(p)+\frac{\partial}{\partial\theta^{2}(p)}\theta^{1}(p)-\frac{\partial}{\partial\theta^{3}(p)}\theta^{4}(p)-\frac{\partial}{\partial\theta^{4}(p)}\theta^{3}(p),*]\,,\\
\hat{\gamma}^{24} & =\int d^{3}p\,[\frac{\partial}{\partial\theta^{1}(p)}\theta^{2}(p)-\frac{\partial}{\partial\theta^{2}(p)}\theta^{1}(p)-\frac{\partial}{\partial\theta^{3}(p)}\theta^{4}(p)+\frac{\partial}{\partial\theta^{4}(p)}\theta^{3}(p),*]\,,\\
\hat{\gamma}^{26} & =\int d^{3}p\,[\frac{\partial}{\partial\theta^{3}(p)}\theta^{1}(p)-\frac{\partial}{\partial\theta^{1}(p)}\theta^{3}(p)+\frac{\partial}{\partial\theta^{4}(p)}\theta^{2}(p)-\frac{\partial}{\partial\theta^{2}(p)}\theta^{4}(p),*]\,,\\
\hat{\gamma}^{27} & =i\int d^{3}p\,[\frac{\partial}{\partial\theta^{3}(p)}\theta^{1}(p)+\frac{\partial}{\partial\theta^{1}(p)}\theta^{3}(p)+\frac{\partial}{\partial\theta^{4}(p)}\theta^{2}(p)+\frac{\partial}{\partial\theta^{2}(p)}\theta^{4}(p),*]\,\\
\hat{\gamma}^{34} & =i\int d^{3}p\,[\frac{\partial}{\partial\theta^{1}(p)}\theta^{1}(p)-\frac{\partial}{\partial\theta^{2}(p)}\theta^{2}(p)+\frac{\partial}{\partial\theta^{3}(p)}\theta^{3}(p)-\frac{\partial}{\partial\theta^{4}(p)}\theta^{4}(p),*]\,,\\
\hat{\gamma}^{36} & =i\int d^{3}p\,[\frac{\partial}{\partial\theta^{4}(p)}\theta^{1}(p)+\frac{\partial}{\partial\theta^{1}(p)}\theta^{4}(p)+\frac{\partial}{\partial\theta^{3}(p)}\theta^{2}(p)+\frac{\partial}{\partial\theta^{2}(p)}\theta^{3}(p),*]\,,\\
\hat{\gamma}^{37} & =\int d^{3}p\,[-\frac{\partial}{\partial\theta^{4}(p)}\theta^{1}(p)+\frac{\partial}{\partial\theta^{1}(p)}\theta^{4}(p)-\frac{\partial}{\partial\theta^{3}(p)}\theta^{2}(p)+\frac{\partial}{\partial\theta^{2}(p)}\theta^{3}(p),*]\,,\\
\hat{\gamma}^{46} & =\int d^{3}p\,[\frac{\partial}{\partial\theta^{4}(p)}\theta^{1}(p)-\frac{\partial}{\partial\theta^{1}(p)}\theta^{4}(p)-\frac{\partial}{\partial\theta^{3}(p)}\theta^{2}(p)+\frac{\partial}{\partial\theta^{2}(p)}\theta^{3}(p),*]\,,\\
\hat{\gamma}^{47} & =i\int d^{3}p\,[-\frac{\partial}{\partial\theta^{4}(p)}\theta^{1}(p)-\frac{\partial}{\partial\theta^{1}(p)}\theta^{4}(p)+\frac{\partial}{\partial\theta^{3}(p)}\theta^{2}(p)+\frac{\partial}{\partial\theta^{2}(p)}\theta^{3}(p),*]\,,\\
\hat{\gamma}^{67} & =-i\int d^{3}p\,[\frac{\partial}{\partial\theta^{1}(p)}\theta^{1}(p)+\frac{\partial}{\partial\theta^{2}(p)}\theta^{2}(p)-\frac{\partial}{\partial\theta^{3}(p)}\theta^{3}(p)-\frac{\partial}{\partial\theta^{4}(p)}\theta^{4}(p),*]\,.
\end{split}
\label{eq:x7a}
\end{equation}

Denote the integrands in (\ref{eq:x1})-(\ref{eq:x2}) as $\hat{\gamma}^{a}(p)$
and in (\ref{eq:x7a}) as $\hat{\gamma}^{ab}(p)$ . So, we can rewrite
(\ref{eq:x1})-(\ref{eq:x2}) as 
\begin{equation}
\hat{\gamma}^{a}=\int d^{3}p\,\hat{\gamma}^{a}(p)\,,\label{eq:x2b}
\end{equation}
and (\ref{eq:x7a}) as 
\begin{equation}
\hat{\gamma}^{ab}=\int d^{3}p\,\hat{\gamma}^{ab}(p)\,.\label{eq:x7c}
\end{equation}

\section{Vacuum and discrete analogs of Grassmann densities}

In \cite{key-17}, the author proposed a method for constructing a
state vector of a vacuum. Let's analyze it in more detail. We divide
the momentum space into infinitely small volumes. We introduce operators
\begin{equation}
\begin{split}B_{k}(p_{j})=\frac{1}{\triangle^{3}p_{j}}\intop_{\triangle^{3}p_{j}}d^{3}p & \,b_{k}(p)\,,\\
\bar{B}_{k}(p_{j})=\frac{1}{\triangle^{3}p_{j}}\intop_{\triangle^{3}p_{j}}d^{3}p & \bar{\,b}_{k}(p)\,.
\end{split}
\label{eq:x8}
\end{equation}

At the same time, given (\ref{eq:x5}), 
\begin{equation}
\{\bar{B}_{k}(p_{i}),B_{l}(p_{j})\}=\frac{1}{\triangle^{3}p_{i}\triangle^{3}p_{j}}\intop_{\triangle^{3}p_{i}}d^{3}p\intop_{\triangle^{3}p_{j}}d^{3}p'\{\bar{b}_{k}(p),b_{l}(p')\}=\frac{1}{\triangle^{3}p_{j}}\delta_{j}^{i}\delta_{l}^{k}\,.\label{eq:x9}
\end{equation}

There is no silent summation over the index that numbers discrete
volumes. For example, it does not exist at index $j$ in (\ref{eq:x8})-(\ref{eq:x9}).
For indexes enclosed in triangular brackets (for example, in (\ref{eq:x11})),
there is also no silent summation.

The expression $\frac{1}{\triangle^{3}p_{j}}\delta_{j}^{i}$ in (\ref{eq:x8})-(\ref{eq:x9})
is a discrete analogue of the delta function $\delta(p-p')$.

In addition, due to the anticommutativity of all $b_{k}(p)$ and $b_{l}(p')$
as well as all $\bar{b}_{k}(p)$ and $\bar{b}_{l}(p')$ it is obvious
that 
\begin{equation}
(B_{k}(p_{j}))^{2}=(\bar{B}_{k}(p_{j}))^{2}=0\,.\label{eq:x10}
\end{equation}

We introduce operators 
\begin{equation}
\begin{split}\Psi_{B_{k}j}= & \triangle^{3}p_{j}B_{<k>}(p_{j})\bar{B}_{<k>}(p_{j})\,,\\
\Psi_{V_{j}}= & \Psi_{B_{1}j}\Psi_{B_{2}j}\Psi_{B_{3}j}\Psi_{B_{4}j}\,
\end{split}
\label{eq:x11}
\end{equation}
and determine through them the fermionic vacuum operator $\Psi_{V}$
\begin{equation}
\Psi_{V}=\prod_{j}\Psi_{V_{j}}\,,\label{eq:x12}
\end{equation}
where the product goes over all physically possible values of $j$.
In this case, we will assume that all volumes $\triangle^{3}p_{j}$
are formed by Lorentz rotations from the volume $\triangle^{3}p_{j=0}$
corresponding to $p=0$, and the grid of angles $\omega_{\mu\nu}$
of these rotations is discrete.

Further, it will often be convenient to represent (\ref{eq:x12})
in the form 
\begin{equation}
\Psi_{V}=\Psi_{V_{j}}\Psi_{V_{j}}^{'}\,,\label{eq:x13}
\end{equation}
where 
\begin{equation}
\Psi_{V_{j}}^{'}=\prod_{i\neq j}\Psi_{V_{i}}\,,\label{eq:x13a}
\end{equation}
is the product of factors in (\ref{eq:x12}), independent of $p_{j}$.

Replace in the formulas with participation of $\hat{\gamma}^{a}$
and $\hat{\gamma}^{ab}$ continuous operators $b_{k}(p)$ and $\bar{b}_{k}(p)$
to discrete $B_{k}(p_{j})$ and $\bar{B}_{k}(p_{j})$, and the integral
$\int d^{3}p\,...$ to the sum $\sum_{j}\triangle^{3}p_{j}\,...$
. In this case, all formulas using continuous operators $b_{k}(p)$
and $\bar{b}_{k}(p)$ are replaced by completely similar ones using
discrete ones, with the replacement of the delta function $\delta(p-p')$
by $\frac{1}{\triangle^{3}p_{j}}\delta_{j}^{i}$ , where $p_{i}$
corresponds to $p$, and $p_{j}$ corresponds to $p'$. We will use
for operators $\hat{\gamma}^{a}=\sum_{j}\triangle^{3}p_{j}\hat{\gamma}^{a}(p_{j})$
and $\hat{\gamma}^{ab}=\sum_{j}\triangle^{3}p_{j}\hat{\gamma}^{ab}(p_{j})$
after such a replacement the same notation as for the corresponding
continuous ones, and we will call such $\hat{\gamma}^{a}$ as discrete
gamma operators, and $\hat{\gamma}^{ab}$ as discrete gamma operators
of rotations.

\section{Action of gamma operators on the vacuum}

Consider action of $\hat{\gamma}^{0}$ on the vacuum (\ref{eq:x13}).
Since $\hat{\gamma}^{0}$ is a commutator, we have 
\begin{equation}
\hat{\gamma}^{0}\Psi_{V}=\hat{\gamma}^{0}\prod_{j}\Psi_{V_{j}}=(\hat{\gamma}^{0}\Psi_{V_{0}})\Psi_{V_{1}}\Psi_{V_{2}}\ldots+\Psi_{V_{0}}(\hat{\gamma}^{0}\Psi_{V_{1}})\Psi_{V_{2}}\ldots+\Psi_{V_{0}}\Psi_{V_{1}}(\hat{\gamma}^{0}\Psi_{V_{2}})\ldots+\ldots\,,\label{eq:x14}
\end{equation}

Here brackets limit the scope of the commutator $\hat{\gamma}^{0}$
. In this case, from (\ref{eq:x11}) it follows 
\begin{equation}
\begin{split}\Psi_{V_{j}}= & (\triangle^{3}p_{j})^{4}B_{1}(p_{j})\bar{B}_{1}(p_{j})B_{2}(p_{j})\bar{B}_{2}(p_{j})B_{3}(p_{j})\bar{B}_{3}(p_{j})B_{4}(p_{j})\bar{B}_{4}(p_{j})\,\end{split}
\label{eq:x15}
\end{equation}

Taking into account the introduced notation for discrete operators
and taking into account the fact that an arbitrary spatial momentum
can be obtained from the state with $p=0$ (\ref{eq:x4}), 
\begin{equation}
\begin{split}B_{1}(p_{j})= & e^{\hat{\gamma}^{0k}\omega_{0k}/2}B_{1}(0)\,,\end{split}
\label{eq:x16}
\end{equation}

At the same time $B_{1}(p_{j})$ means that the result of rotation
of a state with $p=0$ turns into the state with $p=p_{j}$.

First consider action of $\hat{\gamma}^{0}(0)$ on . Easy to see that
\begin{equation}
\begin{split}\hat{\gamma}^{0}\Psi_{V}= & (\hat{\gamma}^{0}\Psi_{V_{0}})\Psi_{V_{1}}\Psi_{V_{2}}\ldots=\\
 & \triangle^{3}p_{j}[\frac{\partial}{\partial\theta^{k}(0)}\theta^{k}(0),\frac{\partial}{\partial\theta^{1}(0)}\theta^{1}(0)\frac{\partial}{\partial\theta^{2}(0)}\theta^{2}(0)\frac{\partial}{\partial\theta^{3}(0)}\theta^{3}(0)\frac{\partial}{\partial\theta^{4}(0)}\theta^{4}(0)]\Psi_{V_{1}}\Psi_{V_{2}}\ldots=0
\end{split}
\label{eq:x16a}
\end{equation}

Now consider action of $\hat{\gamma}^{0}(p)$ on $\Psi_{V}$ for the
case when continious momentum $p=p1$with corresponding discrete $p_{j}$,
that is, it is directed along the axis $\hat{\gamma}^{1}$. Let us
present $\Psi_{V}$ as a product $\Psi_{V}=\Psi_{V_{1,4}}\Psi_{V_{2,3}}\Psi_{V}^{'}$
where 
\begin{equation}
\begin{split}\Psi_{V_{1,4}}= & (\triangle^{3}p_{j})^{2}B_{1}(p_{j})\bar{B}_{1}(p_{j})B_{4}(p_{j})\bar{B}_{4}(p_{j})\,\\
\Psi_{V_{2,3}}= & (\triangle^{3}p_{j})^{2}B_{2}(p_{j})\bar{B}_{2}(p_{j})B_{3}(p_{j})\bar{B}_{3}(p_{j})\,
\end{split}
\label{eq:x17}
\end{equation}

Obviously, 
\begin{equation}
\begin{split}\hat{\gamma}^{0}(p1)\Psi_{V}= & ((\hat{\gamma}^{0}(p1)\Psi_{V_{1,4}})\Psi_{V_{2,3}}+\Psi_{V_{1,4}}(\hat{\gamma}^{0}(p1)\Psi_{V_{2,3}}))\Psi_{V}^{'}\,.\end{split}
\label{eq:x17a}
\end{equation}

Write useful relationships 
\begin{equation}
\begin{split}\frac{\partial}{\partial\theta^{<a>}(p_{j})}\theta^{b}(p_{j})\frac{\partial}{\partial\theta^{<a>}(p_{j})}= & (\frac{1}{\triangle^{3}p_{j}}\delta_{a}^{b}-\theta^{b}(p_{j})\frac{\partial}{\partial\theta^{<a>}(p_{j})})\frac{\partial}{\partial\theta^{<a>}(p_{j})}=\frac{1}{\triangle^{3}p_{j}}\delta_{a}^{b}\frac{\partial}{\partial\theta^{<a>}(p_{j})}\,,\\
\theta^{<a>}(p_{j})\frac{\partial}{\partial\theta^{b}(p_{j})}\theta^{<a>}(p_{j})= & (\frac{1}{\triangle^{3}p_{j}}\delta_{a}^{b}-\frac{\partial}{\partial\theta^{b}(p_{j})}\theta^{<a>}(p_{j}))\theta^{<a>}(p_{j})=\frac{1}{\triangle^{3}p_{j}}\delta_{a}^{b}\theta^{<a>}(p_{j})\,.
\end{split}
\label{eq:x18}
\end{equation}

Consider action of $\hat{\gamma}^{0}(p1)$ on $\Psi_{V_{1,4}}$ and
$\Psi_{V_{2,3}}$. From (\ref{eq:x16}), (\ref{eq:x7a}) and (\ref{eq:x17}),
taking into account (\ref{eq:x18}), we obtain with $p_{j}=p1$ 
\begin{equation}
\begin{split}\hat{\gamma}^{0}(p1)\Psi_{V_{1,4}} & =\frac{\triangle^{3}p_{j}}{2}\sinh\omega_{01}(p1)(\frac{\partial}{\partial\theta^{4}(p1)}\frac{\partial}{\partial\theta^{1}(p1)}+\theta^{1}(p1)\theta^{4}(p1))\,,\\
\hat{\gamma}^{0}(p1)\Psi_{V_{2,3}} & =\frac{\triangle^{3}p_{j}}{2}\sinh\omega_{01}(p1)(\frac{\partial}{\partial\theta^{3}(p1)}\frac{\partial}{\partial\theta^{2}(p1)}+\theta^{2}(p1)\theta^{3}(p1))\,.
\end{split}
\label{eq:x19}
\end{equation}

To understand the meaning of (\ref{eq:x19}) we consider the action
of the operator of creation of a fermion-antifermion pair $\triangle^{3}p\bar{B_{1}}(p)\bar{B}_{4}(p)\approx\triangle^{3}p\theta^{1}(p1)\theta^{4}(p1)$
on $\Psi_{V_{1,4}}$ when $p\rightarrow0$. The multiplier $\triangle^{3}p$
is necessary for normalization to the unit probability of finding
spinors in the whole space.

That is, $\hat{\gamma}^{0}(p1)\Psi_{V_{1,4}}$ contains a term corresponding
to the creation of a fermion-antifermion pair $\theta^{1}(p1)\theta^{4}(p1)$,
suppressed by a small multiplier $\sinh\omega_{01}(p1)$ in the non-relativistic
limit. And $\hat{\gamma}^{0}(p1)\Psi_{V_{2,3}}$ corresponds to the
creation of a pair $\theta^{2}(p1)\theta^{3}(p1)$ with different
values of the spin.

Similarly, $\frac{\partial}{\partial\theta^{4}(p1)}\frac{\partial}{\partial\theta^{1}(p1)}$
are the creation operators of a fermion-antifermion pair for an alternative
vacuum \cite{key-14}, where factors $\frac{\partial}{\partial\theta^{1}(p1)}\theta^{1}(p1)\frac{\partial}{\partial\theta^{4}(p1)}\theta^{4}(p1)$
in the vacuum state vector are replaced by $\theta^{1}(p1)\frac{\partial}{\partial\theta^{1}(p1)}\theta^{4}(p1)\frac{\partial}{\partial\theta^{4}(p1)}$
, and similarly for operator $\frac{\partial}{\partial\theta^{3}(p1)}\frac{\partial}{\partial\theta^{2}(p1)}$
for a corresponding alternative vacuum.

So, $\hat{\gamma}^{0}(p1)\Psi_{V}\rightarrow0$ when $p1\rightarrow0$.

Carrying out the spatial rotations $exp(\hat{\gamma}^{kl}\omega_{kl}/4)$,
where$k,l=1,2,3$, of expressions (\ref{eq:x19}) that do not affect
the multiplier $\hat{\gamma}^{0}$, since $\hat{\gamma}^{kl}$ commutes
with $\hat{\gamma}^{0}$, we get a similar result when for arbitrary
directions of the spatial momentum . Thus, in the non-relativistic
limit $p\rightarrow0$ can be considered $\hat{\gamma}^{0}\Psi_{V}=0$.

Similarly, we find the result of the action of $\hat{\gamma}^{1}(0)$
on the multipliers of $\Psi_{V}$: 
\begin{equation}
\begin{split}\hat{\gamma}^{1}(0)\Psi_{V_{1,4}} & =(\triangle^{3}p_{j})^{3}(\frac{\partial}{\partial\theta^{4}(0)}\frac{\partial}{\partial\theta^{1}(0)}+\theta^{1}(0)\theta^{4}(0))\,,\\
\hat{\gamma}^{1}(0)\Psi_{V_{2,3}} & =(\triangle^{3}p_{j})^{3}(\frac{\partial}{\partial\theta^{3}(0)}\frac{\partial}{\partial\theta^{2}(0)}+\theta^{2}(0)\theta^{3}(0))\,.
\end{split}
\label{eq:x20}
\end{equation}

That means the creation of fermion-antifermion pairs by the operator
$\hat{\gamma}^{1}$ even at zero momentum, that is, without suppression
of this process in the non-relativistic limit.

Thus, the operator $\hat{\gamma}^{1}$ in the nonrelativistic limit
$p\rightarrow0$ (and, therefore, in general) cannot have eigenvalues
on state vectors.

We get the same situation for acting on a vacuum and on state vectors
for operators $\hat{\gamma}^{a}$, $a=1,2,3,4,6,7$ -- they do not
annul the vacuum in the nonrelativistic limit and cannot have eigenvalues
on state vectors.

\section{Action of gamma operators of rotations on a vacuum}

We get the same situation for acting on the vacuum and on state vectors
for boosts $\hat{\gamma}^{0a}(0)$ ,$a=1,2,3,4,6,7$ -- they do not
annul the vacuum and cannot have eigenvalues on the state vectors.

But the rotation operators , $\hat{\gamma}^{kl}(0)$ ,$k,l=1,2,3,4,6,7$,
have the same features as $\hat{\gamma}^{0}(0)$ -- they annul the
vacuum and can have their own eigenvalues ( ) on the state vectors.

The invariance of the vacuum during Lorentz rotations $exp(\hat{\gamma}^{\mu\nu}\omega_{\mu\nu}/4)$,
where$\mu,\nu=0,1,2,3$, is ensured by the fact that each volume $\triangle^{3}p_{j}$
passes into another volume $\triangle^{3}p_{k}$, and its place is
occupied by the third volume $\triangle^{3}p_{l}$. Which only leads
to a change in the order of the factors $\Psi_{V_{j}}$ in (\ref{eq:x12}).
These factors commute, so the Lorentz rotations leave the vacuum $\Psi_{V}$
invariant.

From (\ref{eq:x19}), (\ref{eq:x20}) and similar formulas for all
$\hat{\gamma}^{a}$ and $\hat{\gamma}^{ab}$,$a,b=0,1,2,3,4,6,7$
, it follows that these operators are not Lorentz-invariant (which
is obvious), and their eigenvalues on state vectors can be spoken
only in the non-relativistic limit $p\rightarrow0$.

\section{Lorentz-invariant gamma operators}

It is easy to construct Lorentz-invariant analogs $\hat{\Gamma}^{a}$
and $\hat{\Gamma}^{ab}$ of superalgebraic representations $\hat{\gamma}^{a}$
of Dirac matrices and rotation generators $\hat{\gamma}^{ab}$ . To
do this, it is enough in formulas (\ref{eq:x1})- (\ref{eq:x2}),
(\ref{eq:x7a}) replace all operators $\frac{\partial}{\partial\theta^{k}(p)}$
by $b_{k}(p)$, and operators $\theta^{k}(p)$ by $\bar{b}_{k}(p)$.
For example, 
\begin{align}
\hat{\Gamma}^{0} & =\int d^{3}p\,[b_{1}(p)\bar{b}_{1}(p)+b_{2}(p)\bar{b}_{2}(p)+b_{3}(p)\bar{b}_{3}(p)+b_{4}(p)\bar{b}_{4}(p),*]\,,\label{eq:x21}\\
\hat{\Gamma}^{1} & =\int d^{3}p\,[b_{1}(p)b_{4}(p)-\bar{b}_{4}(p)\bar{b}_{1}(p)+b_{2}(p)b_{3}(p)-\bar{b}_{3}(p)\bar{b}_{2}(p),*]\,,\label{eq:22}\\
\hat{\Gamma}^{67} & =-i\int d^{3}p\,[b_{1}(p)\bar{b}_{1}(p)+b_{2}(p)\bar{b}_{2}(p)-b_{3}(p)\bar{b}_{3}(p)-b_{4}(p)\bar{b}_{4}(p),*]\,,\label{eq:x23}
\end{align}
and so on.

In the discrete version of the theory, in the operators $\hat{\Gamma}^{a}$
and $\hat{\Gamma}^{ab}$, as before, continuous operators $b_{k}(p)$
and $\bar{b}_{k}(p)$ are replaced by discrete $B_{k}(p_{j})$ and
$\bar{B}_{k}(p_{j})$, and integrals $\int d^{3}p\,...$ by sums $\sum_{j}\triangle^{3}p_{j}\,...$.

The operators $\hat{\Gamma}^{a}$ and $\hat{\Gamma}^{ab}$ are constructed
by summing (integrating in the continuous case) over spatial momentums
the results of all possible Lorentz rotations of the operators $\hat{\gamma}^{a}(0)$
and $\hat{\gamma}^{ab}(0)$. As a result of such rotations, $\frac{\partial}{\partial\theta^{k}(0)}$
goes to $b_{k}(p)$, and $\theta^{k}(0)$ to $\bar{b}_{k}(p)$ as
in the field operators, as in $\hat{\gamma}^{a}(0)$ and $\hat{\gamma}^{ab}(0)$.

In contrast to $\hat{\gamma}^{a}$ and $\hat{\gamma}^{ab}$, in the
Lorentz transformations the operators $\hat{\Gamma}^{a}$ and $\hat{\Gamma}^{ab}$
do not change either, since, like for the vacuum, the sum element
for some momentum goes into the sum element for another momentum,
and the sum element for the third momentum takes its place. As a result,
these operators are Lorentz-invariant (and therefore also Lorentz-covariant).
For the same reason, if for some values of $a$ and $b$ the operator
$\hat{\gamma}^{a}(0)$ or $\hat{\gamma}^{ab}(0)$ annuls the vacuum,
then $\hat{\Gamma}^{a}$ or $\hat{\Gamma}^{ab}$ annuls the vacuum,
and if $\hat{\gamma}^{a}(0)$ or $\hat{\gamma}^{ab}(0)$ not annuls
the vacuum, then $\hat{\Gamma}^{a}$ or $\hat{\Gamma}^{ab}$ under
the action on the vacuum do not give zero. And for the same reason,
if $\hat{\gamma}^{a}(0)$ or $\hat{\gamma}^{ab}(0)$ has eigenvalue
for the state with $p=0$, then $\hat{\Gamma}^{a}$ or $\hat{\Gamma}^{ab}$
has corresponding eigenvalue for states with any momentums. That is
why operators $\hat{\Gamma}^{a}$ have the same signature as $\hat{\gamma}^{a}(0)$
and, hence, the same signature as $\hat{\gamma}^{a}$.

Therefore, in quantum relativistic field theory, the eigenvalues of
the operators $\hat{\Gamma}^{a}$ and $\hat{\Gamma}^{ab}$ are meaningful
on the state vectors, and the operators $\hat{\gamma}^{a}$ and $\hat{\gamma}^{ab}$
cannot have eigenvalues at all, since they do not annul the vacuum.
Operators $\hat{\gamma}^{a}$ and $\hat{\gamma}^{ab}$ have eigenvalues
only in the non-relativistic limit $p\rightarrow0$.

Since the commutation relations (\ref{eq:x5}) for $b_{i}(p)$ and
$\bar{b}_{k}(p)$ are the same as for $\frac{\partial}{\partial\theta^{i}(p)}$
and $\theta^{k}(p)$ , the commutation relations for $\hat{\Gamma}^{a}$
and $\hat{\Gamma}^{ab}$ are the same as for $\hat{\gamma}^{a}$ or
$\hat{\gamma}^{ab}$. That is, $\hat{\Gamma}^{\mu}$ are also analogs
of Dirac matrices $\gamma^{\mu},\,\mu=0,\,1,\,2,\,3,\,4$, but $\hat{\Gamma}^{6}$
and $\hat{\Gamma}^{7}$ also expand the set of analogs of Dirac matrices
as $\hat{\gamma}^{6}$ and $\hat{\gamma}^{7}$ .

We introduce the superalgebraic analogues \cite{key-17} of the operators
of the number of particles $\hat{N}_{1},\,\hat{N}_{2}$ and antiparticles
$\hat{N}_{3},\,\hat{N}_{4}$ and the charge operator $\hat{Q}$ in
the method of second quantization: 
\begin{equation}
\begin{split}\hat{N}_{k}(p) & =[\bar{b}_{<k>}(p)\,b_{<k>}(p),*]=-[b_{<k>}(p)\,\bar{b}_{<k>}(p),*]\,,\\
\hat{Q} & =\int d^{3}p\,(\hat{N}_{1}(p)+\hat{N}_{2}(p)-\hat{N}_{3}(p)-\hat{N}_{4}(p))\,,
\end{split}
\label{eq:x24}
\end{equation}

Then the physical meaning of $\hat{\Gamma}^{0}$ and $\hat{\Gamma}^{67}$
is obvious, since (\ref{eq:x21}) and (\ref{eq:x23}) can be rewritten
in the form: 
\begin{equation}
\begin{split}\hat{\Gamma}^{0} & =-\int d^{3}p\,(\hat{N}_{1}(p)+\hat{N}_{2}(p)+\hat{N}_{3}(p)+\hat{N}_{4}(p))\,,\\
\hat{\Gamma}^{67} & =i\int d^{3}p\,(\hat{N}_{1}(p)+\hat{N}_{2}(p)-\hat{N}_{3}(p)-\hat{N}_{4}(p))\,=i\hat{Q}\,.
\end{split}
\label{eq:x25}
\end{equation}

That is, $-\hat{\Gamma}^{0}$ is the operator of the total number
of spinors and antispinors, and $\hat{\Gamma}^{67}$ is related to
the charge operator $\hat{Q}$ by the ratio $\hat{\Gamma}^{67}=i\hat{Q}$.
Similarly, $\hat{\Gamma}^{jk}=i\hat{\tau}_{l}$ , where $j,k,l$ is
cyclic permutation of $1,2,3$. Moreover, $\hat{\tau}_{l}$ are Lorentz-invariant
spin operators, which are analogs of the Pauli matrices. Operators
$\hat{W}{}^{k}=-im\hat{\Gamma}^{k4}$ are components of Lorentz-invariant
analog of spin components of the Pauli-Lyubansky vector. However,
physical meaning of operators $\hat{\Gamma}^{jk}$, $\hat{W}{}^{k}$
and $\hat{\Gamma}^{\mu a}$, where $\mu=0,1,2,3,\,a=4,6,7$, is incomprehensible.

Under the action on the vacuum (\ref{eq:x12})

\begin{equation}
\hat{\Gamma}^{0}\Psi_{V}=\hat{\Gamma}^{mn}\Psi_{V}=0\,,
\end{equation}
where $m,n=1,2,3,4,6,7$, and the other gamma operators do not give
zero under the action on $\Psi_{V}$. Therefore, you can measure only
the eigenvalues of the operators $\hat{\gamma}^{0}(0)$, $\hat{\gamma}^{mn}(0)$,
$\hat{\Gamma}^{0}$ and $\hat{\Gamma}^{mn}$ with $m,n=1,2,3,4,6,7$.

It is useful to note that the matrix formalism does not provide the
possibility of zero eigenvalues of gamma matrices, in contrast to
the proposed theory.

\section{Spacetime signature in the presence of a spinor vacuum}

The reason for the difference between the action on the vacuum and
the state vectors of the operators $\hat{\gamma}^{0}(0)$ and $\hat{\gamma}^{mn}(0)$,
on the one hand, and $\hat{\gamma}^{m}(0)$, $\hat{\gamma}^{0m}(0)$,
on the other, is related to the structure of these operators in (\ref{eq:x1})-(\ref{eq:x2}),
(\ref{eq:x7a}), (\ref{eq:x21})-(\ref{eq:x23}). Since the vacuum
state vector has a multiplier $B_{<i>}(0)\,\bar{B}_{<i>}(0)$, the
action on vacuum of operators consisting only of terms of the form
$[\bar{B}_{l}(0)B_{k}(0),*]$ will always give zero, since, by virtue
of (\ref{eq:x9}) and (\ref{eq:x10})

\begin{equation}
[\bar{B}_{<l>}(0)B_{<k>}(0),\,B_{<k>}(0)\bar{B}_{<k>}(0)B_{<l>}(0)\bar{B}_{<l>}(0)]=0\,.
\end{equation}

But the terms of the form $[B_{k}(0)\,B_{l}(0),*]$ and $[\bar{B}_{k}(0)\,\bar{B}_{l}(0),*]$
will give a non-zero result. Summing the results of Lorentz rotations
leads to similar conclusions for $\hat{\Gamma}^{0}$ , $\hat{\Gamma}^{mn}$,
on the one hand, and $\hat{\Gamma}^{m}$ , $\hat{\Gamma}^{0m}$, on
the other.

Expansion (\ref{eq:x3}) generates the expansion of field operators
in momenta and leads to the implementation of the Dirac equation \cite{key-18}.
The question arises of what kind of Clifford bases such decomposition
is possible.

If, as in the considered case, $\hat{\gamma}^{0}=(\hat{\gamma}^{0})^{+},\,\hat{\gamma}^{m}=-(\hat{\gamma}^{m})^{+}$
, there is one time-like Clifford vector.

Multiplying $\hat{\gamma}^{0}$ by an imaginary unit will lead to
the appearance in the expansion in momenta \cite{key-18} of exponentially
increasing terms, that is, to the impossibility of the existence of
normalized solutions. Therefore, Clifford vectors $\hat{\gamma}^{0}$
and $\hat{\Gamma}^{0}$ are time-like and have signature +1 for spacetime
where spinors can exist as physical particles.

Multiplication of any of the operators $\hat{\gamma}^{m}$ (and, consequently,
$\hat{\Gamma}^{m}$) by the imaginary unit due to the presence of
the vacuum (\ref{eq:x12}) will lead to asymmetry between Clifford
vectors $\hat{\Gamma}^{0}$ and $i\hat{\Gamma}^{m}$, since $\hat{\Gamma}^{0}\Psi_{V}=0$
and $i\hat{\Gamma}^{m}\Psi_{V}\neq0$, and $\hat{\Gamma}^{0}$ can
have eigenvalues on the state vectors but $i\hat{\Gamma}^{m}$ cannot.
The space of Clifford vectors with the same signature must be isotropic,
however in this case we obtain a preferred direction. Therefore, other
than $\hat{\Gamma}^{0}$ Clifford vectors could not have the same
signature as $\hat{\Gamma}^{0}$. Consequently, the condition for
the existence of the vacuum imposes restrictions on the possible variants
of Clifford algebras: neither complex algebra nor algebras in which
at least one of the base vectors $\hat{\Gamma}^{m}$ (and hence $\hat{\gamma}^{m}$
) is timelike is suitable. Therefore, all Clifford vectors $\hat{\Gamma}^{m}$
are spacelike (and hence $\hat{\gamma}^{m}$) -- they have a signature
of -1, and there is only one basic timelike Clifford vector $\hat{\Gamma}^{0}$
(and hence $\hat{\gamma}^{0}$).

Only 16 of the 28 operators $\hat{\Gamma}^{a}$ and $\hat{\Gamma}^{ab}$
in (\ref{eq:x3}) annul the vacuum and therefore can have eigenvalues
on the state vectors. Therefore, if we require the existence of a
decomposition in momenta, that is, the existence of spinors as physical
particles, out of seven gamma matrices $\hat{\Gamma}^{a}$ (and hence
$\hat{\gamma}^{a}$ ), one must have a positive signature, and the
other six must have a negative signature.

Thus, in the superalgebraic theory of spinors, the signature of a
four-dimensional spacetime can only be (1, -1, -1, -1), and there
are two additional axes $\hat{\gamma}^{6}$ and $\hat{\gamma}^{7}$
with a signature (-1, -1) corresponding to the inner space of the
spinor. The reason why they and the axis $\hat{\gamma}^{4}$ are not
additional spatial axes is not yet clear.


\section{Discussion}

The proposed theory has a number of interesting consequences.

-- Expansion (\ref{eq:x3}) ensures for spinors the existence of
decomposition in momenta \cite{key-18}.

-- The theory is free from divergences, leading to the need for the
normalization of operators \cite{key-17}.

-- It leads to an unambiguous signature of spacetime, which coincides
with the observable.

-- Part of the decomposition terms (\ref{eq:x3}) corresponds to
the usual field theories available in the framework of the general
theory of relativity \cite{key-19}, as well as to theories of bundles
\cite{key-20}. An operator $\hat{\Gamma}^{67}=i\hat{Q}$ and gauge
transformation $exp(i\hat{Q}\omega_{67})$ automatically arises, where
$\hat{Q}$ is the charge operator in the second quantization formalism,
$\hat{Q}\Psi=\Psi$ for the spinor $\Psi$, and $\hat{Q}\bar{\Psi}=-\bar{\Psi}$
for its antiparticle $\bar{\Psi}$.

-- The proposed approach to constructing a discrete vacuum is fundamentally
different from theories in which the discreteness of spacetime is
considered, leading to the loss of Lorentz covariance \cite{key-21}.
The proposed theory is Lorentz-covariant and combines the features
of discrete and continuous theories.

-- We can construct operators $\hat{\Gamma}^{a}$ and $\hat{\Gamma}^{ab}$,
$a,b=0,1,2,3,4,6,7$ from operators of creation and annihilation of
spinors independently on superalgebraic representation $\hat{\gamma}^{\mu}$
of Dirac gamma matrices $\gamma^{\mu}$. However this representation
makes interconnection between Dirac gamma matrices and operators $\hat{\Gamma}^{a}$
obvious.


\section{Conclusions}

In this article, we investigated action of operator analogs of Dirac
gamma matrices (we called them gamma operators) on a vacuum.

We derived formulas for vacuum state vector and operators of the Lorentz
transformations of spinors in the superalgebraic representation of
spinors. Five operator analogs of five Dirac gamma matrices exist
in the superalgebraic approach as well as two additional operator
analogs of gamma matrices. Gamma operators are constructed from Grassmann
densities and derivatives with respect to them.

We have shown that the gamma operator $\hat{\gamma}^{0}$ when acting
on a vacuum state vector gives zero in the limit of zero momentum.
The rest of the gamma operators $\hat{\gamma}^{1}$, $\hat{\gamma}^{2}$,
$\hat{\gamma}^{3}$, $\hat{\gamma}^{4}$, $\hat{\gamma}^{6}$, $\hat{\gamma}^{7}$,
when acting on a vacuum, do not give zero.

Similar results are observed for generalized Lorentz rotation operators.
Operators $\hat{\gamma}^{12}$, $\hat{\gamma}^{13}$ and so on, corresponding
to rotations in the planes of Clifford vectors (gamma operators) $\hat{\gamma}^{1}$,
$\hat{\gamma}^{2}$, $\hat{\gamma}^{3}$, $\hat{\gamma}^{4}$, $\hat{\gamma}^{6}$,
$\hat{\gamma}^{7}$, annihilate the vacuum in the limit of zero momentum,
but the boost operators $\hat{\gamma}^{01}$, $\hat{\gamma}^{02}$,
$\hat{\gamma}^{03}$, $\hat{\gamma}^{04}$, $\hat{\gamma}^{06}$,
$\hat{\gamma}^{07}$ do not.

We have shown that there are gamma operators $\hat{\Gamma}^{a}$ and
$\hat{\Gamma}^{ab}$, which are built from creation and annihilation
operators, and that they are also analogs of Dirac gamma matrices.
However, unlike gamma operators $\hat{\gamma}^{a}$ and $\hat{\gamma}^{ab}$,
they are Lorentz-invariant.

Lorentz-invariant gamma operator $\hat{\Gamma}^{0}$ when acting on
a vacuum gives zero without any limit by momentum. The rest of the
gamma operators $\hat{\Gamma}^{1}$, $\hat{\Gamma}^{2}$, $\hat{\Gamma}^{3}$,
$\hat{\Gamma}^{4}$, $\hat{\Gamma}^{6}$, $\hat{\Gamma}^{7}$, when
acting on a vacuum, do not give zero.

That is why only real algebra with one timelike basis Clifford vector
corresponding to the zero gamma matrix in the Dirac representation
can be realized. In this case, the signature of the four-dimensional
spacetime, in which there is a vacuum state, can only be (1, -1, -1,
-1), and there are two additional axes corresponding to the inner
space of the spinor, with a signature (-1, -1).

It is useful to note that the matrix formalism does not provide the
possibility of zero eigenvalues of gamma matrices, in contrast to
the proposed theory.

So, we have shown that the condition for the existence of spinor vacuum
imposes restrictions on possible variants of Clifford algebras of
gamma operators.

\end{document}